# The Methicillin-Resistant Staphylococcus Aureus Infection Controls

A Review

Jiapu Zhang[*1]

[1]Graduate School of Sciences, IT and Engineering, CIAO, The University of Ballarat, Mt Helen, Ballarat, Vic 3353, Australia

[*1]Correspondence address: email j.zhang@ballarat.edu.au, jiapu_zhang@hotmail.com, telephone 61-3-5327 6335, 61-4 2348 7360

*Abstract*

*Background:* Multi-resistant organisms (MROs), the bacteria that are resistant to a number of different antibiotics, have been very popular around the world in recent years. They are very difficult to treat but highly infectious in humans. MRSA (Methicillin-Resistant Staphylococcus Aureus) is one of the MROs. It is believed that in 2007 more people died of MRSA than of AIDS worldwide. In Australia "there are about 2000 people per year who have a bloodstream infection with the MRSA germ and the vast majority of those get them from health care procedure" (Nader, 2005). It is acknowledged as a significant challenge to Australian hospitals for MRSA infection control. Nursing professionals are in urgent need of the study of MRSA nosocomial infection controls.

*Rationale:* This review provides insight into the hand washing and isolation infection-control strategies for MRSA. Electronic searches were undertaken of Medine, CINAHL, Health source: Nursing/Academic, internurse.com, Meditext and Google Advanced Search. Primary search words and phrases such as infection control, Multi-resistant organisms, MRSA, hand hygiene, hand wash, alcohol based handrub, isolation and environment control were used. The important technologies on those two aspects worldwide are well surveyed, compared, contrasted, and discussed.

*Purpose:* The review is to do a complete survey on the hand washing and isolation technologies of infection controls for MRSA and try to provide some possible recommendations for Australian hospitals.

## Introduction

In Australia antimicrobial resistance is increasing in many pathogens and has occurred in health care acquired infections. Examples occurred include methicillin resistant Staphylococcus aureus (MRSA), vancomycin resistant enterococci (VRE), extended-spectrum beta-lactamases (ESBLs), and multi-resistant acinetobacter baumannii (MRAB), etc. These examples are called multi-resistant organisms (MROs), because they are resistant to a number of antimicrobial drugs and thus the drugs cannot not kill the organism any more. Australia has a history of high antibiotic usage. As we all know, Australia is the original country of the development of penicillin (as well as its variations) for use as a medicine. In Australia when we go to see the doctor, the doctor always prescribe the medicine of penicillin's variations. Anti-microbial use in the emergence of MROs is a major concern in Australia and it may be due to overuse, misuse, and inappropriate prescribing.

In Victoria, since the late 1970s MRSA strains have been identified as a major cause of hospital acquired infections. MRSA accounts for approximately 30 to 50 percent of hospital acquired Staphylococcus aureus (S. aureus). MRSA is a strain of S. aureus that is resistant to all beta-lactam antibiotics, for example penicillins. It is a Gram-positive bacterium commonly found on human skin and in the noses of healthy people and it can enter the body causing infection, which may be sometimes fatal (blood or wound infections). The example of burn wound infections is the three MRSA patients transferred to Royal Perth Hospital after October 2002 Bali bombings (Heath et al, 2003, Silla et al, 2006). MRSA has become more serious than before. According to website for MESA from Wikipedia, we get to know in 2007 "more people died of MRSA than of AIDS worldwide". In Australia "there are about 2000 people per year who have a bloodstream infection with the MRSA germ and the vast majority of those get them from health care procedure" (Nader, 2005). It is really a significant challenge to Australian hospitals for MRSA infection control.

Two effective strategies for MRSA infection controls are alcohol-based and the isolation. Alcohol-based strategy is to use alcohol for surface sanitizing and use alcohol-based hand rubs for hand washing. Alcohol plus quaternary ammonium have been proven to be a long effective surface sanitizer against MRSA. NAV-CO2 is used to effectively against MRSA pathogens in hospitals, ambulances, and nursing homes. Alcohol-based hand rubs are proved to be a successful hand washing strategy. It was reported by the Centers for Disease Control and Prevention (CDC) only by hand washing strategy every year 30,000 patients are saved from nosocomial infections. Screening and isolation seem to be another effective strategy for infection-control. All patients with MRSA were immediately isolated, and all staff were screened for MRSA and were prevented from working until they had completed a course of eradication therapy that was proven to work. Dutch succeeded for this strategy in attempting eradication of carriage upon discharge from hospital for the MRSA colonised patients. If colonised patients are discharged back into the community and then readmitted, the loss of control will occur; US and UK have been overwhelmed by MRSA in this way and they have put more investments on facilities.

Working with MRSA patients, nursing professionals are in urgent to study nosocomial infection controls. This essay does a complete survey on the hand washing and isolation technologies of infection controls and provides insight into the hand washing and isolation infection-control strategies for MRSA. Some possible recommendations are presented for Australian hospitals.

**The infection control strategy of alcohol-based handrubs**

Alcohol-based hand rub (ABHR) is recommended as the primary choice for hand decontamination by CDC and the clinic validation can be found in paper (Gordin et al, 2005). The clinic impact was observed in an inner-city tertiary-care teaching hospital of Washington. An observational survey for 6 years comparing the first 3 years of no ABHR use with the following 3 years of using ABHR was done: in the first 5 years and more an antimicrobial soap with 0.3% triclosan was provided for staff hand hygiene, in the next 3 years the wall-mounted dispensers of an ABHR with 62.5% ethyl alcohol placed in all inpatient and all outpatient clinic rooms was used, the data of the last 6 years were collected. On comparison of the first 3 years with the final 3 years, there was a 21% decrease in new nosocomially acquired MRSA (90 to 71 isolates per year; P=0.01). However, for Clostridium difficile the incidence was essentially unchanged (Gordin et al, 2005, Vernaz et al, 2008) though it was increased by the Sterillium ABHR technology (King, 2004). Paper (Vernaz et al 2008) differs from paper (Gordin et al, 2005) in that the ABHRs were analyzed combined with the aggregated data on antibiotic use, the observation time is 6 years and 7 months, and the observation was not done at single hospital, instead of at several Geneva University Hospitals. With the use of different antibiotic classes, the consumption of ABHR is increased for MRSA incidence (Vernaz et al, 2008). The limitations of (Vernza et al, 2008) were the lack of detailed data on the number of admission cultures, as pointed out by (Harbarth & Samore, 2008) which also discussed the MRSA incidence of one or more antibiotic drug classes usage. Widmer (2007) introduced practical education on proper technique for using ABHR tested by the addition of a fluorescent dye, which can significantly increase the degree of bacterial killing. To compare the effect of ABHR with other hand hygiene product, Larson et al (2005) also test a traditional antiseptic handwash and found that nurses' skin condition was improved using ABHR. However, a new alcohol-based hand hygiene product is alcohol gel, which is not only reduces the number of inpatients newly affected by MRSA but also the antibiotic costs (MacDonald et al, 2004). Alcohol gel was first used in a 600-bedded district general hospital of UK for cleaning hands between clinical contact with patients. The case notes of patients newly affected by nosocomial MRSA were reviewed for 1 year before and after the performance feedback of hand hygiene, and the cost of teicoplanin use (for MRSA infections) was also determined for those two periods. A significant reduction in the number of inpatients newly affected by MRSA (P<0.05) and in the use of teicoplanin was observed. This shows that alcohol gel can reduce nosocomial MRSA infection rate and antibiotic use. The above-mentioned papers (Gordin et al, 2005, Vernaz et al, 2008, MacDonald et al, 2004) stand for the most important advances on effective alcohol-based infection control strategy for hand decontamination.

## Isolation strategy for infection control

Screening patients for MRSA and isolating MRSA-positive patients is the so-called isolation strategy for control of nosocomial MRSA infection. Nationwide "search-and-destroy" technology (van Trijp et al, 2007) to control MRSA colonization and infection was effective but controversial. There are a lot of other technologies, such as on polymerase chain reaction screening (Cunningham et al, 2007, Huletsky et al, 2005) and active surveillance cultures (Shitrit, 2006, Boyce et al, 2004), for this strategy, along with many factors considered such as the implications both in cost and resources, the effect of MRSA infection on patient morbidity and mortality, and the need for individual risk assessment of MRSA colonized or infected patients to prevent the transmission of MRSA isolates (Bissett, 2005). Bissett (2005) wrote a systematic survey to discuss screening and isolating patients to control the MRSA infection and pointed out that screening and isolation should be universal to decelerate the rate of transmission of MRSA. The theory, reasons, development and need of antimicrobial resistance of MRSA, and government response to infection control are also introduced in (Bissett, 2005). The inactivation of penicillin by Staphylococcus aureus-produced beta-lactamase is well illustrated in Figure 1 of (Bissett, 2005). (Bissett, 2005) is an excellent paper that I picked up on isolation and screening. The screen and isolation strategy of (Bissett, 2005) is discussed as follows.

Routine screening within primary care settings is not effective, especially for elder people. In high-risk areas, such as intensive care unit (ICU), all patients and nurses need to be screened until the MRSA status of all colonized patients is known. The screening in high-risk areas can be more costly to individual and government than control. Single-room source isolation is effective to reduce the risk of transmission of MRSA. When considering the use of source isolation, the good factors that require assessment are: knowledge of the colonizing microorganisms, how the microorganism is transmitted, the site of the body affected, whether patients or staff are vulnerable, and the measures required for containment, which all are clearly listed in Table 2 of (Bissett, 2005). MRSA can survive in dry and dusty environments and can be spread in the airborne route or contact. Thus, to limit the area of contamination and the appropriate use of gloves and efficient hand hygiene are also the isolation from MRSA colonized or infected patients. The risk of isolation is needed to assess when the source isolation is used. Many good practice measures and their corresponding rationales for isolation and screening can be found in Tables 3 and 1 of (Bissett, 2005).

To explain the research findings of (Bissett, 2005) further more, the paper (Cepeda et al, 2005) is used. Single-room accommodation usage in some sense reminds staff and visitors that special precautions are required and indicates that the movement of staff between patients and facilities make the spread of microorganisms. (Cepeda et al, 2005) aims to assess the effectiveness of moving versus not moving infected or colonised patients, who are isolated in single rooms or cohorts, in ICUs to reduce spread of MRSA, because single room or cohort isolation benefit or risk over or above other contact precautions is not known. The prospective 1-year study was carried out in 3 general medical-surgical ICUs of 2 central London teaching hospitals. Admissions and weekly screenings were used to find the incidence of MRSA colonization. The 1-year is divided into 3 phases: the first 3 months (move phase), the middle 6 months (non-move phase), and the last 3 months (move phase). In phase 2, MRSA-positive patients were not moved to a single room or cohort nursed unless they were carrying other multi-resistant or notifiable pathogens. At the end of each phase, existing MRSA patients were treated as new admissions. Throughout the 1-year, standard and contact precautions were practiced, hand hygiene was encouraged, and compliance was audited. At last the findings of (Bissett, 2005) cannot confirm that isolation of ICU patients who are colonized or infected with MRSA over and above the use of standard precautions in an endemic environment. This might be due to the prospective 1-year of period is not enough for the observations.

## Recommendations

Basing on the above discussions, some basic recommendations are given for MRSA infection controls. Firstly, every practitioner should recognize the precaution of MRSA. Secondly, after admissions, each patient should accept an active surveillance culture of MRSA to identify whether the patient should access isolation practice. Before getting the result of test, the patient should be treated as an isolate. In addition, all staff need to be screened for MRSA and prevented from working until they have completed a course of eradication therapy that was proven to work. Thirdly, before and after each procedure, hand washing is necessary. On the other hand, hospitals should provide ABHRs or alcohol gel to reduce nosocomial transmission of MRSA. Last but not least, the optimal choice for isolation of MRSA is single room. If hospitalization condition is limited by financial supports and hospital infrastructure, cohorts of patients with MRSA infection can be used.

## Conclusions

Nowadays hospitals around the world face with the challenge how to prevent and control the infections of MROs. MRSA is one of the most important MROs. This essay specially reviewed and discussed its strategies on infection controls. The infection control strategy of ABHRs and isolation strategy for infection control are two basic and most important MRSA infection control strategies. Both are reviewed and discussed in details. At last some recommendations are given. Sometimes the two basic known strategies do not work ideally; new strategies for MRSA infection control is still challenging to health associated workers.

**REFERENCES**

Boyce, J.M., Havill, N.L., Kohan, C., Dumigan, D.G., and Ligi, C.E. "Do infection control measures work for methicillin-resistant staphylococcus aureus?" Infection Control and Hospital Epidemiology 25 (2004): 395–401.

Cepeda, J.A., Whitehouse, T., Cooper, B., Hails, J., Jones, K., Kwaku, F., Taylor, L., Hayman, S., Cookson, B., Shaw, S., Kibbler, C., Singer, M., Bellignan, G., and Wilson, A.P. "Isolation of patients in single rooms or cohorts to reduce spread of MRSA in intensive-care units: prospective two-centre study." Lancet 365(9456) (2005): 295-304.

Cunningham, R., Jenks, P., Northwood, J., Wallis, M., Ferguson, S., and Hunt, S. "Effect on MRSA transmission of rapid PCR testing of patients admitted to critical care." Journal of Hospital Infection 65(1) (2007): 24-28.

Gordin, F.M., Schultz, M.E., Huber, R.A., and Gill, J.A. "Reduction in nosocomial transmission of drug-resistant bacteria after introduction of an alcohol-based handrub." Infection Control and Hospital Epidemiology 26(7) (2005): 650-653.

Harbarth, S., and Samore, M.H. "Interventions to control MRSA: high time for time-series analysis?" Journal of Antimicrobial Chemotherapy 62(3) (2008): 431-433.

Heath, C.H., Orrell, T.C., Lee, R.C., Pearman, J.W., McCullough C. and Christiansen, K.J. "A review of the Royal Perth Hospital Bali experience: an infection control perspective." Australian Infection Control 8(2) (2003): 43-54.

Huletsky, A., Lebel, P., Picard, F.J., Bernier, M., Gognon, M., Boucher, N., and Bergeron, M.G. "Identification of methicillin-resistant staphylococcus aureus." Clinical Infectious Diseases 40 (2005): 976-981.

King, S. "Provision of alcohol hand rub at the hospital bedside: a case study." Journal of Hospital Infection 56 (2004): 10-12.

Larson, E.L., Cimiotti, J., Haas, J., Parides, M., Nesin, M., Della-Latta, P., and Saiman, L. "Effect of antiseptic handwashing vs alcohol sanitizer on health care-associated infections in neonatal intensive care units." Archives of Pediatrics and Adolescent Medicine 159(4) (2005): 377-383.

MacDonald, A. "Performance feedback of hand hygiene, using alcohol gel as the skin decontaminant, reduces the number of inpatients newly affected by MRSA and antibiotic costs." Journal of Hospital Infection 56(1) (2004): 56-63.

Nader, C. "Hospital superbug found in 120 deaths." The Age, August 17, 2005. Accessed May 29, 2013. http://www.theage.com.au/news/national/hospital-superbug-found-in-120-deaths/2005/08/16/1123958064556.html

Silla, R., Fong, J., Wright, J., and Wood, F. "Infection in acute burn wounds following the Bali bombing: a comparative prospective audit.' Burns 32(2) (2006): 139-144.

"Methicillin-resistant Staphylococcus aureus", from Wikipedia, the free encyclopedia. Accessed May 29, 2013. http://en.wikipedia.org/wiki/Methicillin-resistant_Staphylococcus_aureus

Shitrit, P., Gottesman, B.-S., Katzir, M., Kilman, A., Ben-Nissan, Y., and Chowers, M. "Active surveillance for methicillin-resistant staphylococcus aureus (MRSA) decreases the incidence of MRSA bacteremia." Infection Control and Hospital Epidemiology 27 (2006): 1004-1008.

van Trijp, M.J.C.A., Melles, D.C., Hendriks, W.D.H., Parlevliet, G.A., Gommans, M., and Ott, A. "Successful control of widespread methicillin-resistant staphylococcus aureus colonization and infection in a large teaching hospital in the Netherlands." Infection Control and Hospital Epidemiology 28 (2007): 970-975.

Vernaz, N., Sax, H., Pittet, D., Bonnabry, P., Schrenzel, J., and Harbarth, S. "Temporal effects of antibiotic use and hand rub consumption on the incidence of MRSA and Clostridium difficile." Journal of Antimicrobial Chemotherapy 62(3) (2008): 601-607.

Widmer, A.F., Conzelmann, M., Tomic, M., Frei, R., and Stranden, A.M. "Introducing alcohol-based hand rub for hand hygiene: the critical need for training." Infection Control on Hospital Epidemiology 28(1) (2007): 50-54.